\documentclass[aps,prl,reprint,showpacs]{revtex4-1}

\pdfoutput=1
\usepackage{color}
\usepackage{graphicx}
\usepackage{amsmath}
\usepackage{amssymb}

\begin{document}
\title{From Coulomb blockade to nonlinear quantum dynamics\\
in a superconducting circuit with a resonator}
\date{\today}
\author{Vera Gramich}
\email{vera.gramich@uni-ulm.de}
\affiliation{Institut f\"{u}r Theoretische Physik, Universit\"{a}t Ulm,
  Albert Einstein-Allee 11, 89069 Ulm,
  Germany}
\author{Bj\"orn Kubala}
\thanks{V.~Gramich and B.~Kubala contributed equally to this work.}
\affiliation{Institut f\"{u}r Theoretische Physik, Universit\"{a}t Ulm,
  Albert Einstein-Allee 11, 89069 Ulm,
  Germany}
  \author{Selina Rohrer}
\affiliation{Institut f\"{u}r Theoretische Physik, Universit\"{a}t Ulm,
  Albert Einstein-Allee 11, 89069 Ulm,
  Germany}
\author{Joachim Ankerhold}
\affiliation{Institut f\"{u}r Theoretische Physik, Universit\"{a}t Ulm,
  Albert Einstein-Allee 11, 89069 Ulm,
  Germany}

\begin{abstract}
Motivated by recent experiments on superconducting circuits consisting of a dc-voltage biased Josephson junction in series with a resonator, quantum properties of these devices far from equilibrium are studied. This includes a crossover from a domain of incoherent to a domain of coherent Cooper pair tunneling, where the circuit realizes a driven nonlinear oscillator. Equivalently, weak photon-charge coupling turns into strong correlations captured by a single degree of freedom. Radiated photons offer a new tool to monitor charge flow and current noise gives access to nonlinear dynamics, which allows to analyze quantum-classical boundaries.
\end{abstract}

\maketitle
\paragraph{Introduction.-}
The essence of quantum mechanics lies in the existence of pairs of conjugated, non-commuting variables linked by uncertainty relations. For a superconducting Josephson junction (JJ) the nonlinear dynamics of the number of Cooper pairs (CPs) transferred across the junction and the phase difference of the superconducting condensates are linked in that manner \cite{barone:1982,likharev:1986}.
In a dc-biased circuit, the electromagnetic properties of the equilibrium environment and the coupling of the JJ determine, whether a `classical' ac-Josephson current with a well defined phase, an incoherent tunneling of single CPs due to dynamical Coulomb blockade (DCB), or the full quantum dynamic regime, where neither phase nor charge behaves like a classical variable, occurs \cite{grabert:1998}.

In the last years, this set-up has been extended by combining JJ devices with superconducting resonators, the electromagnetic modes of which act as dynamical degrees of freedom. This has led to an unprecedented control of quantum properties such as the creation of cat-like states \cite{hofheinz:2009} and the observation as well as theoretical description of nonlinear dynamics \cite{siddiqi:2005,peano:2006, marthaler:2007, ong:2011,wilson:2011,dykman:2012,ong:2013}. While in these circuits no net charge flows through the JJ,
 dc-voltage biased set-ups, implemented very recently in \cite{basset:2010,hofheinz:2011, chen:2011, pashkin:2011}, offer new possibilities to study nonlinear quantum properties  in a tunable photon-charge system {\em far from equilibrium} \cite{nazarov:2012, leppakangas:2013, blencowe:2012}.

In this Letter we consider a circuit, realized experimentally in \cite{hofheinz:2011},  where a JJ, biased by a voltage $V$, is placed in series to a resonator as displayed in Fig.~\ref{fig:diagram}. At low temperatures and voltages below the superconducting gap, the excess energy $2eV$ of tunneling CPs is completely transformed into photons exciting the resonator.  There, photon leakage leads to radiation that can be detected.
Two limiting scenarios are then possible: Either CP tunneling is slow compared to photon relaxation so that between subsequent tunneling events the cavity returns to its ground state, or charge transfer is fast so that photons accumulate and back-act on the JJ giving rise to strong charge-photon correlations. The first regime, known as DCB, has been analyzed in \cite{hofheinz:2011} and corresponds to an {\em incoherent} CP flow. However, what happens in the second regime and how the crossover between the two domains occurs, is not known yet. The goal of this work is to fill this gap and to provide detailed predictions for future experiments. As we will show, in the second regime, the JJ behaves according to the classical Josephson relations, based on {\em coherent} CP tunneling, and the circuit realizes a driven nonlinear oscillator governed by a single degree of freedom. By tuning its parameters, one may continuously switch between the two domains and thus access different dynamical properties. In a broader context, these results contribute to current efforts to achieve a deeper understanding of quantum--classical boundaries in non-equilibrium systems including superconducting \cite{leppakangas:2013,armour:2013}, micromechanical \cite{sillan:2013}, and cold atom set-ups \cite{schmied:2012}.
\begin{figure}[hb]
\includegraphics[width=0.9\linewidth]{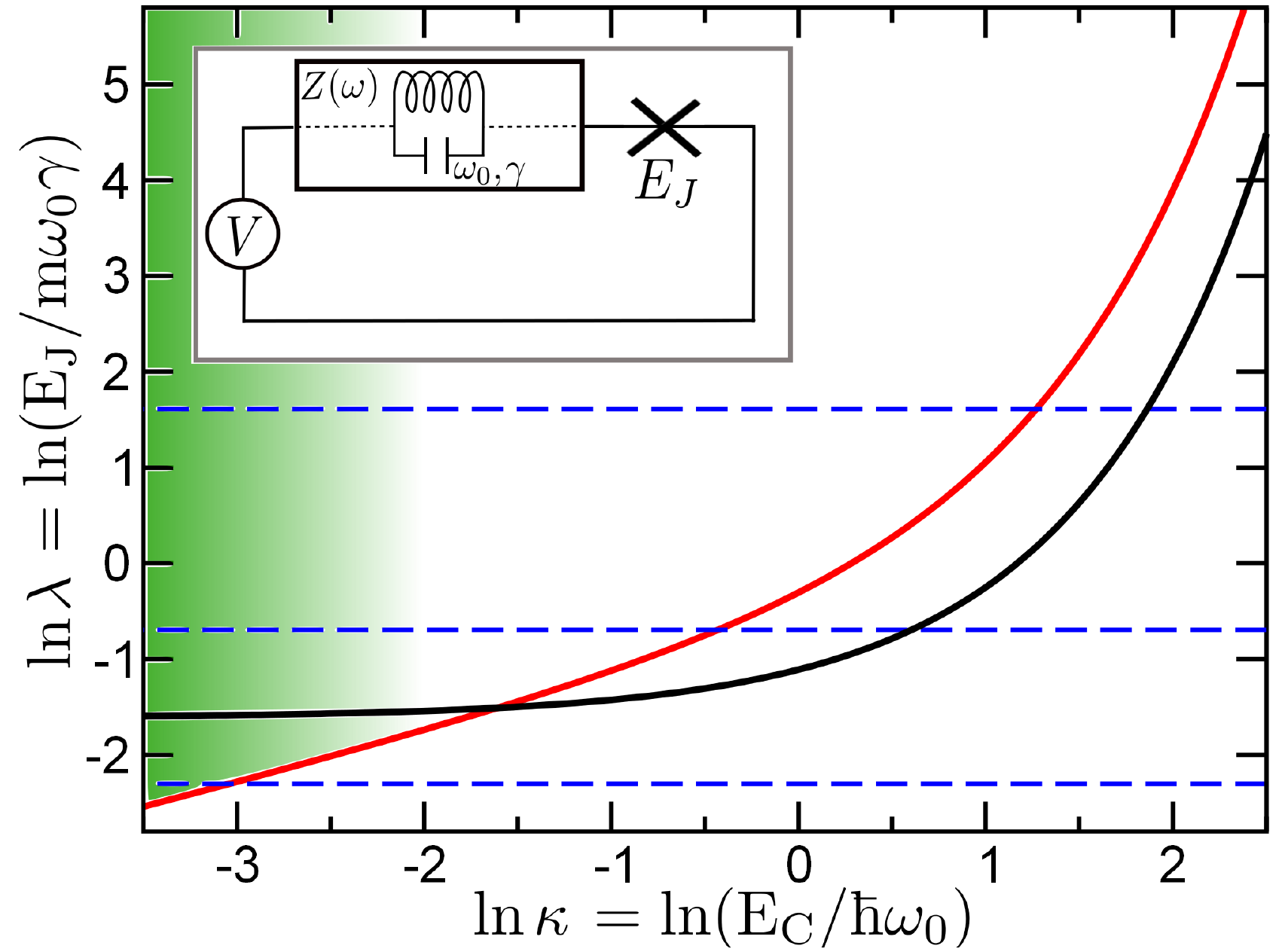}
\caption{\label{fig:diagram} (Color online) Parameter space of a circuit consisting of a resonator with mode frequency $\omega_0$ and damping $\gamma$ in series with a voltage biased JJ (inset) with dimensionless coupling $\lambda =  E_J/(m \omega_0 \gamma)$ and  the scale for DCB $\kappa  = E_C/\hbar\omega_0$: Classical behavior is seen in the green range; below (above)  the red line charge flow through the JJ occurs incoherently (coherently) with the resonator being  at $T=0$ basically empty (excited); the black line separates the domains of linear  (below) and nonlinear (above) dynamics. Horizontal lines (blue) refer to the parameter values of Fig.~\ref{fig:limit_test}.}
\end{figure}

\paragraph{Model.-}The Hamiltonian of the circuit follows from the two sub-units, the resonator part and the JJ part, where in the regime we are interested in only a single mode of the resonator impedance $Z(\omega)$ is relevant. One thus has $\tilde{H}_0=\frac{q^2}{2 C}+ \bigl(\frac{\hbar}{2e}\bigr)^2 \frac{1}{2L}\, \phi^2-E_J \cos(\eta)-\, 2 e (V-V_{\rm res}) \,N$ describing a harmonic oscillator with mass $m=\bigl(\frac{\hbar}{2e}\bigr)^2 C$ and frequency $\omega_0=1/\sqrt{LC}$ in series to a JJ with phase $\eta$ which is subject to a bias voltage $V$. Here, the two sets of conjugate variables obey $[q, \phi]=-2 i e$ and $[N, \eta]=-i$, where in contrast to the charge operator $q$, the number operator $N$ has discrete eigenvalues, counting the number of CPs that have transferred the JJ. The effective voltage at the JJ also contains the voltage drop at the resonator $2 e V_{\rm res}=-\hbar\dot{\phi}$, thus coupling JJ and resonator dynamically.

Now, a straightforward calculation shows \cite{epaps} that for weak detuning $\Delta/\omega_0=(\omega_0-\omega_J)/\omega_0\ll1 $ with $\omega_J=2eV/\hbar$, in the rotating frame the Hamiltonian $\tilde{H}_0$ takes the form
\begin{equation}\label{hamilton}
{H}_0=\hbar\Delta a^\dagger a+i\frac{E_J^*}{2} : \left( a^\dagger \, {\rm e}^{i\eta}-  a\, {\rm e}^{-i\eta}\right) \frac{J_1(2\sqrt{\kappa\, n})}{\sqrt{n}}:\,
\end{equation}
with standard photon annihilation/creation operators of the resonator and ${\rm e}^{\pm i\eta}$ invoking forward/backward CP tunneling. Here, $:\ :$ denotes normal ordering and the Bessel function $J_1$ of the first kind contains the photon number operator $n=a^\dagger a$. The dimensionless parameter
\[
\kappa=E_C/\hbar\omega_0\equiv \hbar/(2 m\omega_0)
 \]
 represents the scale for charge quantization through $E_C=2e^2/C$ while $E_J^*=E_J\, {\rm e}^{-\kappa/2}$ is a renormalized Josephson energy \cite{grabert:1998}.

According to the experimental setting \cite{hofheinz:2011}, the electromagnetic environment of the circuit consists of high frequency modes acting as a heat bath and low frequency voltage noise. The former leads to photon leakage from the resonator while the latter one can be seen as a fluctuating component of the bias voltage which thus couples to the charge $N$. At low temperatures,
 the dynamics of the reduced density operator of the JJ-resonator compound is then captured by a master equation
 \begin{equation}\label{master}
\dot{\rho}=-\frac{i}{\hbar}[H_0,\rho]+ \frac{\gamma}{2} {\cal{L}}[a, \rho] +\frac{\gamma_J}{2} {\cal{L}}[N, \rho]\, ,
\end{equation}
where dissipators ${\cal{L}}[x,\rho]$ model the impact of the respective environments. The rate $\gamma$ determines the photon lifetime in the cavity via its $Q$-factor, i.e.\ $Q=\omega_0/\gamma$, and the rate $\gamma_J$ follows from the noise power of low frequency voltage fluctuations, see \cite{epaps}.  This decreases with decreasing temperature so that $\gamma_J\ll \gamma$. Hence, we start by putting $\gamma_J=0$ and discuss further details below.

The dynamics in (\ref{master}) displays a complex interplay between charge transfer and photon emission/absorption. Restricting ourselves to the stationary state, this is particularly seen in the resonator population which according to Eq.~(\ref{master})  reads
\begin{equation}\label{occupation_Bessel}
\langle n\rangle_{\rm st} = \frac{ \lambda\,{\rm e}^{-\frac{\kappa}{2}}}{4 \kappa} \langle : \left(a^\dagger  \, {\rm e}^{i\eta}+  a\, {\rm e}^{-i\eta}\right) \frac{J_1(2\sqrt{\kappa\, n})}{\sqrt{n}}:\rangle_{\rm st}\, .
\end{equation}
 Here, we introduced the coupling parameter
 \[
 \lambda=E_J/(m\omega_0\gamma)
 \]
 which, as shown below, determines together with the quantum parameter $\kappa$ the dynamics of the circuit. The current through the JJ, i.e.\ $\langle I_J\rangle_{\rm st}\equiv 2 e \langle \dot{N}\rangle_{\rm st}$, is obtained accordingly from (\ref{master}) and turns out as $\langle I_J\rangle_{\rm st}=2e\gamma \langle n\rangle_{\rm st}$. This reflects energy conservation between charge flow and photon absorption. Photon radiation in steady state is, of course, also fixed by $\langle n\rangle_{\rm st}$. Correlations of charges and photons are captured by $\langle a {\rm e}^{-i\eta}\rangle_{\rm st}= C_{a, \eta}\, \gamma/(\gamma-2 i\Delta)$ with
\begin{equation}\label{correlator}
 C_{a, \eta}= \frac{\lambda\,{\rm e}^{-\frac{\kappa}{2}}}{2\sqrt{\kappa}}\, \langle :\!  J_0(2\sqrt{\kappa\, n}) + \frac{a}{a^\dagger}\, J_2(2\sqrt{\kappa\, n})\, {\rm e}^{-2 i\eta}\!:\rangle_{\rm st}\, .
\end{equation}
While in general explicit results must be obtained numerically, it is intriguing to first consider how limiting cases are recovered and what are their precise ranges of validity.

\paragraph{Incoherent charge transfer.-}
In the weak coupling regime  $\lambda\ll 1$ (cf.~Fig.~\ref{fig:diagram}), the Bessel functions can be linearized by assuming self-consistently $\kappa \langle n\rangle_{\rm st}\ll 1$. One then arrives with (\ref{hamilton})  at the Hamiltonian for DCB \cite{ingold:2000}, i.e.\ $H_{0, \rm DCB}= \hbar\Delta n +(i/2) \sqrt{\kappa} E_J^* ( a^\dagger \, {\rm e}^{i\eta}-  a\, {\rm e}^{-i\eta})$. Likewise,  we find for the photon occupation (\ref{occupation_Bessel})  with $C_{a, \eta}\approx \lambda{\rm e}^{-\kappa/2}/2\sqrt{\kappa}$ in leading order
\begin{equation}\label{n_P(E)}
\langle n\rangle^\textrm{lin}_{\rm st}  = \frac{\lambda^2 {\rm e}^{-\kappa}}{4\, \kappa} \,\frac{\gamma^2}{\gamma^2+4\Delta^2}\, ,
\end{equation}
which in turn verifies the assumption. The current $\langle I_J\rangle^\textrm{lin}_{\rm st}=2e\gamma \langle n\rangle^\textrm{lin}_{\rm st}$ is identical to the one derived within the golden rule treatment of DCB [$P(E)$-theory] \cite{ingold:2000, hofheinz:2011} describing {\em incoherent} CP transport across the JJ. Subsequent tunneling processes  are thus
statistically independent and, in the low temperature range, one may express the current also in terms of the forward tunneling rate $\Gamma_{\rm f}$ as $\langle I_J\rangle_{\rm st}=2 e \Gamma_{\rm f}$  with $\Gamma_{\rm f}=\gamma  \langle n\rangle^\textrm{lin}_{\rm st}$. This implies that between subsequent tunneling events the resonator returns to its ground state (at $T=0$) and there is no back-action onto the JJ. The condition for this scenario is that photon relaxation occurs sufficiently fast compared to CP tunneling, i.e.\ $\Gamma_{\rm f}\ll \gamma \Rightarrow \langle n\rangle^\textrm{lin}_{\rm st} \ll1$. This relation defines the crossover between incoherent and coherent charge flow (red line in Fig.~\ref{fig:diagram}): For fixed $\kappa$, the coherent domain is approached by increasing $\lambda$ and thus by either increasing the Josephson energy or the photon lifetime in the cavity.

We note that formally the linearization condition used in (\ref{n_P(E)}) $\kappa \langle n\rangle_{\rm st}\ll 1$ is not identical to the condition for the incoherent-coherent transition $\langle n\rangle_{\rm st} \ll1$,  (cf.~Fig.~\ref{fig:diagram}). Physically, for small $\kappa$ and $\lambda$ the circuit may thus display linear dynamics even in the coherent regime.

\paragraph{Classical regime.-}
 We now consider the situation where the photon occupation and in turn the JJ current are  large, while the quantum parameter $\kappa\ll 1$ such that $\kappa \langle n\rangle_{\rm st}= {\rm const.}$ Charging effects thus do not play any role and we may put ${\rm e}^{i\eta}\to 1$ and replace operators by $2\sqrt{\kappa} a\to  Z \exp(i \varphi)$ with real-valued amplitude $Z$ and phase $\varphi$. Consequently, at resonance ($\Delta=0$) the rotating frame  Hamiltonian (\ref{hamilton}) reduces to
\begin{equation}\label{hamiltoncl}
H_{0,\rm cl}= E_J J_1(Z)\, \sin(\varphi)\, .
\end{equation}
The same result is also obtained directly from a classical description of the circuit in Fig.~\ref{fig:diagram}: The Kirchhoff rules impose $V=V_J+V_{\rm res}$ so that the voltage $V_J$ across the JJ is slaved to the dynamics of the resonator phase $V_{\rm res}=-(\hbar/2e)\dot{\phi}$.  Accordingly, the classical Josephson energy reads $-E_J \cos(\phi+\omega_J t)$
and acts as a nonlinear drive on the resonator. Near resonance $\omega_J\approx \omega_0$, the ansatz $\phi(t)=Z \cos(\omega_J t+\varphi)$  for the stationary orbit then leads in the rotating frame to (\ref{hamiltoncl}). We emphasize that in contrast to most driven nonlinear oscillators, recently realized also with superconducting circuits (see e.g.\ \cite{dykman:2012}), here, the nonlinearity is part of the drive and not part of a static potential. The classical treatment is based on the {\em coherent} flow of CPs and the circuit is described by the single degree of freedom $\phi$ of a driven nonlinear system. Amplitudes and phases of stationary orbits are determined by the static parts of the classical equations of motions \cite{epaps}, i.e.,
\begin{equation}\label{classical}
Z^2 = 2 \lambda\, J_1(Z) \cos(\varphi)\ \ , \ \  0= \sin(\varphi)\, dJ_1(Z)/dZ \, .
\end{equation}
For $\lambda\ll 1$, this set of equations has only one solution, namely, the  orbit $Z_0\approx \lambda, \varphi_0=0$ obtained by linearizing $J_1(Z)$. With increasing $\lambda$, the amplitude grows, nonlinearities become relevant, and at $\lambda=\lambda_c\approx 3$ a first bifurcation occurs. There, a new class of orbits appears with constant amplitude $Z_1\approx 1.8$, given by $[dJ_1(Z)/dZ](Z_1)=0$, and growing phase $\varphi_1(\lambda>\lambda_c)>0$. The  critical coupling $\lambda_c$ follows from $Z_0(\lambda_c)=Z_1$.

\begin{figure}[hb]
\includegraphics[width=0.98\linewidth]{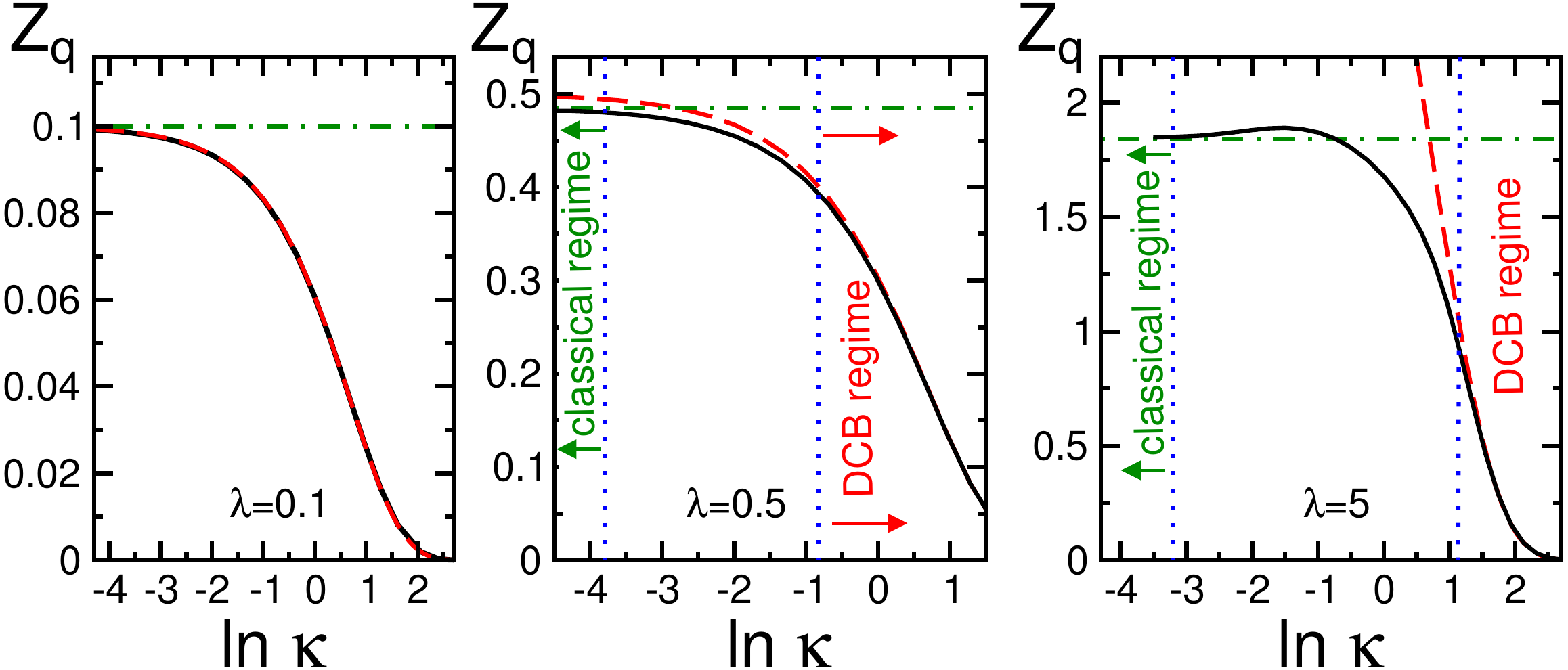}
\caption{\label{fig:limit_test} (Color online) Quantum amplitude $Z_{\rm q}=2\sqrt{\kappa\langle n\rangle_{\rm st}}$ vs.\ $\kappa$ for different couplings $\lambda$. Green (red) lines show the classical (DCB), black lines the full nonlinear quantum results. Corresponding ranges of validity are indicated (cf.~Fig.~\ref{fig:diagram}).}
\end{figure}

\paragraph{Full Quantum dynamics.-}
The master equation, Eq.~(\ref{master}), describes the full dynamics of the JJ-resonator system. The density $\rho$ is conveniently found numerically in a base of product states that are eigenstates of the number operators $n$ and $N$.
However, in this basis $\rho(t)$  does {\em not} reach a stationary state due to a finite current $\langle I_J\rangle_{\rm st}\sim {\rm Tr}\{N\dot{\rho}(t)\}$. This problem is circumvented by introducing auxiliary densities $ \rho_\chi= \textrm{Tr}_J\{ e^{ i \chi \eta} \rho\}$ with a partial trace over the JJ degrees of freedom and $\chi$ being an integer. Coherences between differing numbers of transferred Cooper-pairs are captured for $\chi\neq 0$. This way, one arrives  at a hierarchy of coupled equations of motions for the $ \rho_\chi$ which is solved by proper truncation. Based on the quantum regression theorem \cite{carmichael:2002}, all relevant observables of cavity and JJ are then evaluated.

\paragraph{Results.-}We are now able to investigate how the crossover from the limiting regimes to the full nonlinear quantum case in Fig.~\ref{fig:diagram} is encoded in various observables which are accessible experimentally. \\
The fact that the resonator occupation determines in the classical limit the amplitude of the orbit via $\hbar\omega_0\langle n\rangle_{\rm st}\to m\omega_0^2 Z^2/2$, suggests to formally define $Z_{\rm q}=2\sqrt{\kappa\langle n\rangle_{\rm st}}$. We may then study how for fixed $\lambda$ this `quantum amplitude' approaches the classical domain for small $\kappa$ and the DCB domain for large $\kappa$, see Fig.~\ref{fig:limit_test}. In case of $\lambda\ll 1$, the DCB-result (\ref{n_P(E)}) provides a fairly accurate description over the full range where quantum effects only appear in the renormalized parameter $\lambda {\rm e}^{-\kappa/2}$. In contrast, for larger couplings nonlinearities are relevant in the classical regime (smaller $\kappa$) as well as in the non-perturbative quantum domain where CP transfer is coherent (intermediate values of $\kappa$). These results verify the domains in parameter space depicted in Fig.~\ref{fig:diagram}. Moreover, we recover from the simulations the  experimental observations of Ref.~[\onlinecite{hofheinz:2011}] obtained in the DCB regime (see  Fig.~\ref{fig:P(E)_results}) \footnote{In the actual experiment finite temperature corrections are relevant as well as a more extended form of the resonator impedance. $P(E)$-theory assumes an equilibrium environment of the JJ which for $T=0$ corresponds to an empty resonator}. Since $\langle I_J\rangle_{\rm st} \propto \langle n\rangle_{\rm st}$ and $\omega_J$ varies with the voltage $V$, the occupation $\langle n\rangle_{\rm st}$ as a function of the detuning (cf.~Fig.~\ref{fig:P(E)_results} right) provides also the $IV$-curve of the JJ.

 Experimentally of particular relevance are photon correlations such as $g^{(1)}(\tau)=\langle a^{\dagger}(t)a(t+\tau)\rangle/\langle a^{\dagger}a\rangle$. Its Fourier transform provides for long times $t$ the spectral distribution of the photon radiation. We find that its width sensitively depends on the low frequency voltage fluctuations which have been neglected so far. Upon comparing experimental data with numerical predictions for $\gamma_J\neq 0$, one gains $\gamma_J /\gamma\approx 0.04 \ll 1$, verifying that they are relevant only for those quantities which are broadened solely by $\gamma_J$. The next order correlation $g^{(2)}(\tau)=\langle a^{\dagger}(t)a^{\dagger}(t+\tau)a(t+\tau)a(t)\rangle/\langle a^{\dagger}a\rangle^2$ carries information about correlations between emitted photons and thus indicates (anti-)bunching, see Fig.~\ref{fig:P(E)_results}. Even though charge flow is incoherent in the DCB regime, $g^{(2)}(\tau)$ reveals non-Poissonian photon correlations $g^{(2)}(\tau)\neq 1$. For weak driving $\lambda \ll 1$, $g^{(2)}(0)$ is related to the probability of finding the resonator excited by a second tunneling event, before it relaxes to its ground state, with the result $g^{(2)}(0)=(1-\kappa/2)^2$.
\begin{figure}[ht]
\includegraphics[width=0.95\linewidth]{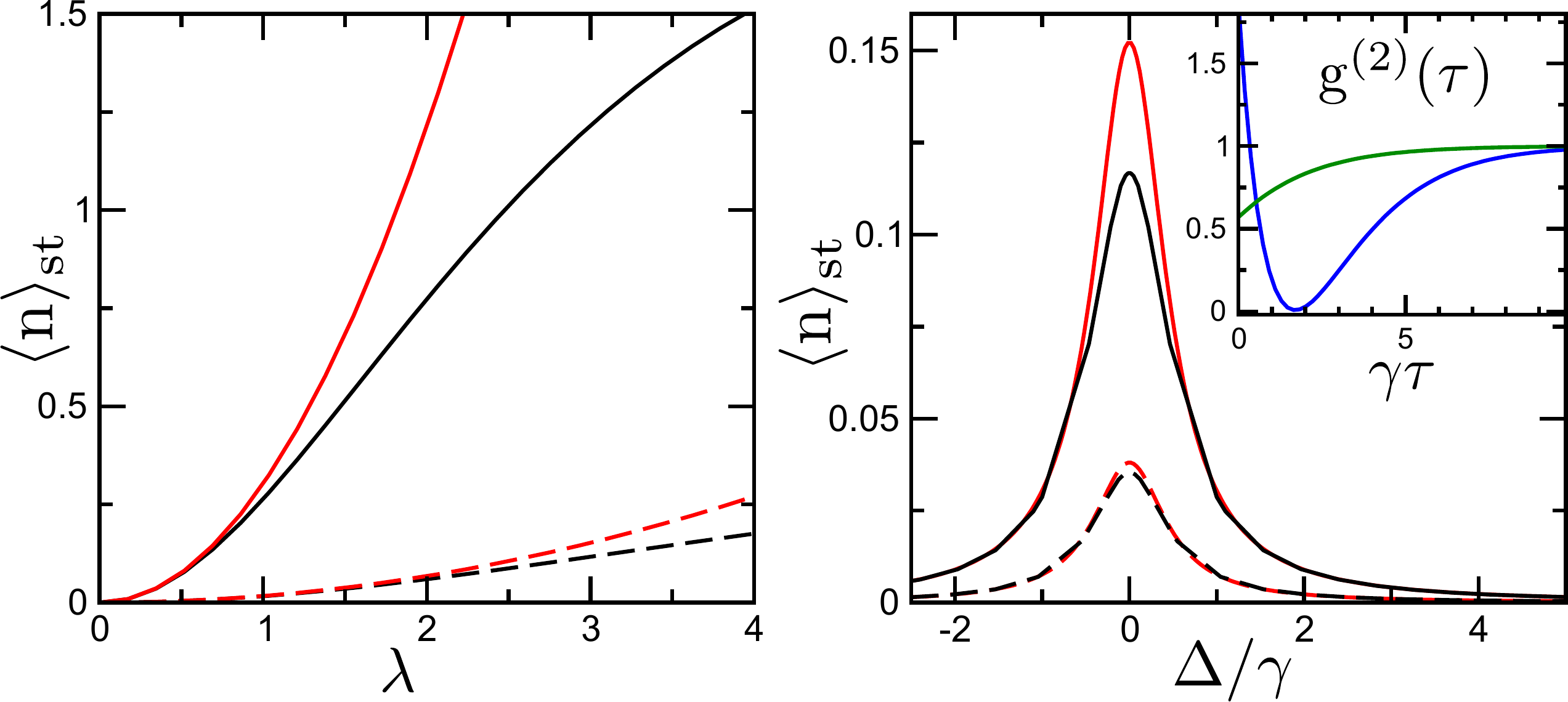}
\caption{\label{fig:P(E)_results} (Color online) Photon occupation $\langle n \rangle_\textrm{st}$ (black) vs. coupling $\lambda$ (left) and detuning $\Delta$ (right). Exact results (black) are shown together with the DCB result (\ref{n_P(E)}) (red). Left: $\kappa=0.5$ (solid), $\kappa=2$ (dashed). Right:  $\lambda=1.5$ (dashed),  $\lambda=3$ (solid). Inset: Green (blue) lines reveal (anti)bunching for $\kappa=0.5$ ($\kappa=4.7$) visible in $g^{(2)}(\tau)$ for $\lambda\ll 1$.}
\end{figure}

 The circuit considered here provides not only a new tool to analyze  charge flow by detecting emitted photons, but also to monitor nonlinear dynamics by detecting current correlations as e.g.\ the Fano factor $F_J=S_{I_JI_J}/(4e I_J)$ of the JJ-current-current noise $S_{I_JI_J}$, see Fig.~\ref{fig:Fano}. As expected, we find $F_J=1$ in the incoherent, single-CP transport regime (shot noise) for weak coupling. The onset of a coherent charge flow through the JJ results in a substantial drop of $F_J$ due to a reduction of shot noise. However, most strikingly the Fano factor approaches a minimum followed by a pronounced peak exactly at those values for $\lambda$ where according to (\ref{classical}) new classical orbits emerge. As this class of new orbits is a mere consequence of the nonlinearity and exists even in absence of dissipation (formally $\lambda\to \infty$), the resonator gains on average (over one driving period) no net energy from the driving source. Physically, this means that a new channel for a correlated two-CP transfer opens: The energy quantum $2eV=\hbar\omega_0$ deposited in the cavity by a forward CP transfer is used to promote a backward transfer leading in turn to no net current. Around the classical bifurcation point the competition between two sets of classical orbits is then observable as a substantial increase in the charge noise. For larger $\kappa$, the bifurcation point is shifted according to $\lambda\to \lambda {\rm e}^{-\kappa/2}$ and features are smeared out by quantum fluctuations (Fig.~\ref{fig:Fano}, left). These findings open  fascinating avenues to study signatures of classical bifurcations in the deep quantum regime experimentally.
\begin{figure}[ht]
\includegraphics[width=0.97\linewidth]{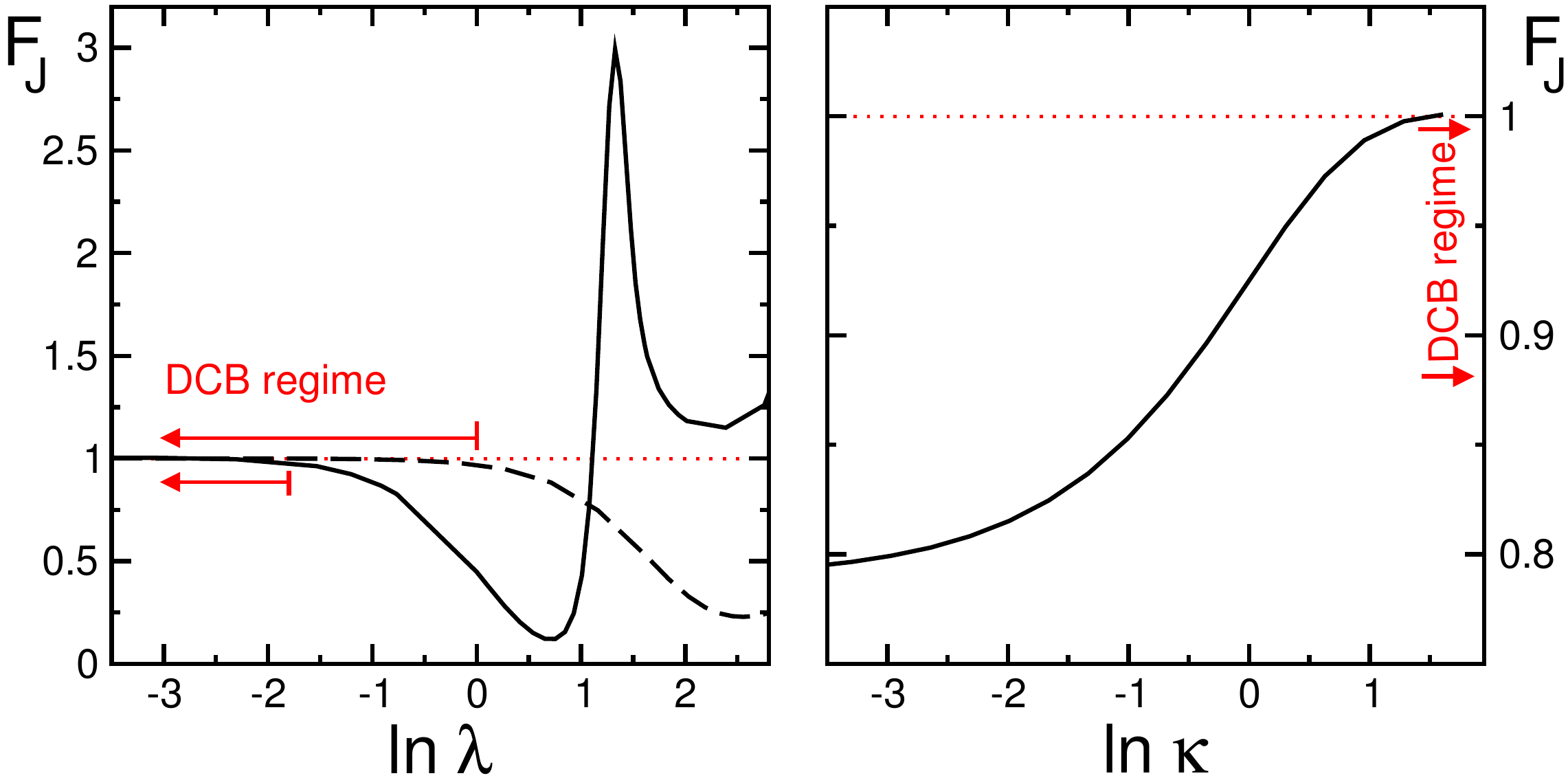}
\caption{\label{fig:Fano} (Color online) Left: Fano factor $F_J$ for the current-current noise vs. $\lambda$ for $\kappa=0.05$ (solid) and $\kappa=3$ (dashed). Right: $F_J$ vs. $\kappa$ for $\lambda=0.5$; incoherent single-CP transfer corresponds to $F_J\equiv 1$ (dots).}
\end{figure}

We conclude this analysis by highlighting that rich physics is also present {\emph{beyond}} the fundamental resonance $\omega_J\approx \omega_0$. For this purpose, one relaxes the rotating-wave approximation in the numerical approach described above which then gives access to further resonances in the resonator occupation (respectively the JJ current) when the applied dc-voltage is varied \footnote{Rotating-wave approximations valid for and close to a particular higher-order resonance can also be employed.}. As illustrated in Fig.~\ref{fig:2_phot}, resonances occur indeed for $\omega_J=p\, \omega_0,\;p\in \mathbb{Z}$. In generalization of (\ref{n_P(E)}), within the DCB-regime one shows that the $p$-th resonance scales as ${\rm e}^{-\kappa}\,\kappa^{|p|-2}/|p|!$. As displayed for $p=2$ in Fig.~\ref{fig:2_phot}, the generation of $p$ photons by a single CP leads to strong correlations in the photon output and thus to a strongly non-Poissonian resonator occupation, see Fig.~\ref{fig:2_phot}. Accordingly, for weak driving the correlation $g^{(2)}(0)=1/(2\langle n \rangle_{\rm st})\gg 1$ diverges (cf.~Ref.~[\onlinecite{leppakangas:2013}]).
\begin{figure}[h]
\hspace*{0.03\columnwidth}
\includegraphics[width=0.88\columnwidth]{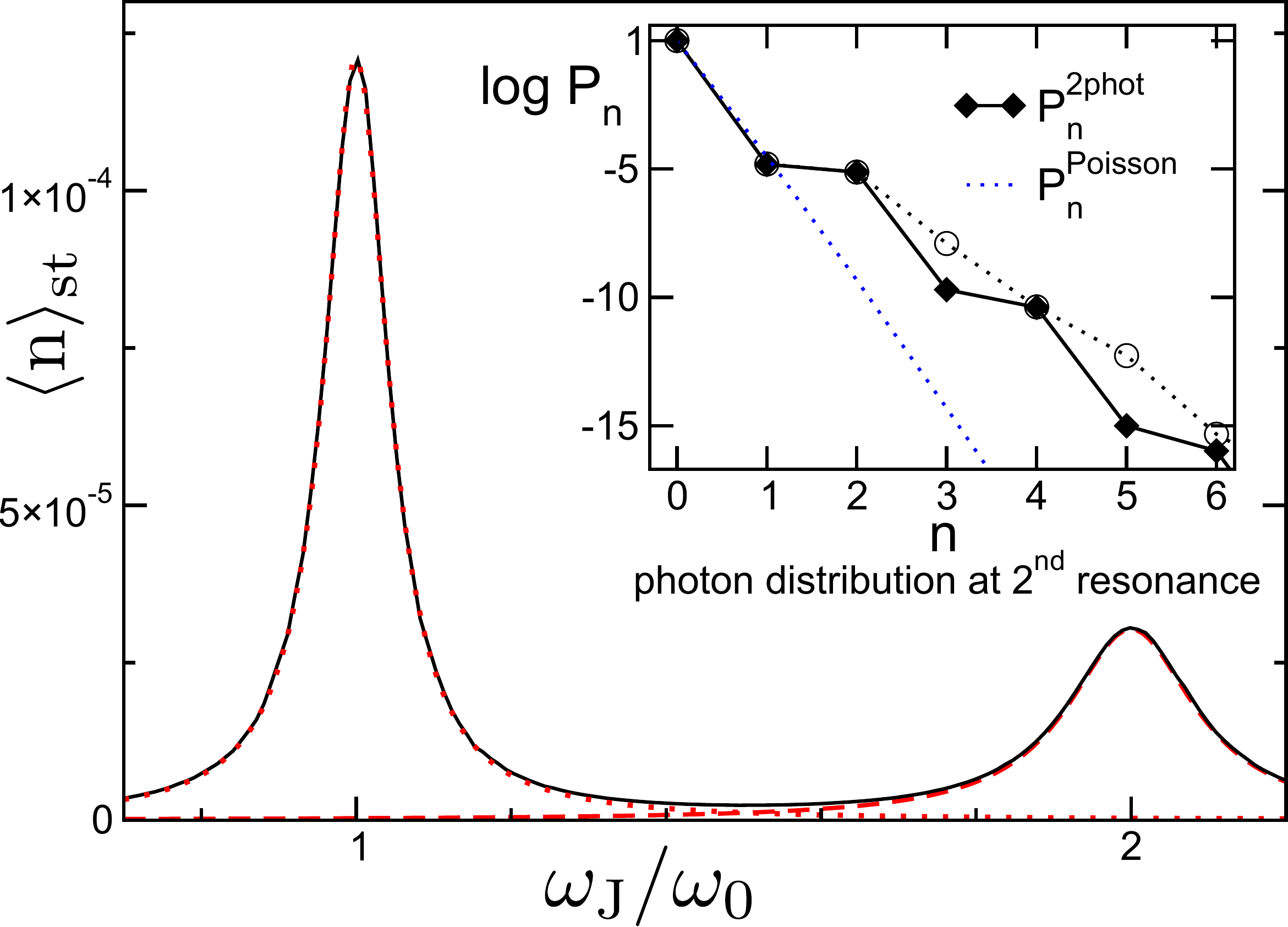}
\caption{\label{fig:2_phot} (Color online) One and two photon resonances in the cavity occupation  for $\kappa=0.5, \lambda=0.02, Q=10$ (black) together with DCB results (red). Two photon processes lead to a non-Poissonian photon distribution (inset) for $Q=10$ (circles) and $Q\rightarrow \infty$ (diamonds).}
\end{figure}

To summarize, we have analyzed the quantum dynamics of a superconducting circuit consisting of a voltage biased JJ in series with a resonator in strong non-equilibrium. Analytical findings and numerical simulations provide detailed information about the crossover from the regime of sequential tunneling with weak photon-charge correlations to the one where the circuit behaves as a driven nonlinear oscillator with a single degree of freedom. This also implies a quantum-classical transition.
Charge flow is detectable via photon radiation and current-current correlations display quantum signatures of classical bifurcations in the nonlinear regime.
Multi-photon resonances reveal complex charge-photon interaction including photon bunching and are thus of great interest for future theoretical and experimental studies.

\paragraph{Acknowledgements.} The authors thank A. Armour, M. Blencowe, M. Hofheinz, F. Portier,  and A. Rimberg for valuable discussions. JA, BK, and VG like to thank for the kind hospitality of the Department of Physics and Astronomy, Dartmouth College, Hanover, NH, (JA, BK) and the O.V. Lounasmaa Laboratory, Aalto University, Helsinki, Finland, (VG).
Financial support was provided by Deutsche Forschungsgemeinschaft
through AN336/6-1 and SFB/TRR21. VG also acknowledges gratefully the German Academic Exchange Service (DAAD) and the European Community's Seventh Framework Programme (FP7/2007-2013) under
grant agreement No. 228464 (MICROKELVIN). SR acknowledges the support of the Zeiss Foundation.

\bibliographystyle{apsrev4-1}

%

\begin{widetext}
\appendix
\section{Supplemental Material}

\begin{small}
\noindent
In this supplemental material to our article `From Coulomb blockade to nonlinear quantum dynamics
 in a superconducting circuit with a resonator' we present further details about the derivation of the circuit Hamiltonian in the rotating frame, the description
of the sources for decoherence in the master equation, and the classical equations of motion.
\end{small}

\maketitle

\section{Hamiltonian of the circuit and open quantum dynamics}
The circuit analyzed in the main text is described by a Hamiltonian of the form $H=\tilde{H}_0+\tilde{H_I}+\tilde{H}_R$. Here,
$\tilde{H}_0=H_{\rm res}+H_J-2 e V_J N$ captures the resonator with impedance $Z(\omega)$ in series with a Josephson junction (JJ) which is biased by an external voltage $V$ so that the effective voltage for the CP transfer at the JJ is $V_J=V-V_{\rm res}$. For the further discussion only a single resonator mode with phase $\phi$ is taken into account so that $H_{\rm res}=\frac{q^2}{2 C}+ \bigl(\frac{\hbar}{2e}\bigr)^2 \frac{1}{2L}\, \phi^2$ with $[q, \phi]=-2i e$. Physically, this is equivalent to a harmonic oscillator with frequency $\omega_0=1/\sqrt{LC}$ and mass $m=(\hbar/2e)^2 C$.
The JJ part is given by the standard expression $H_J=-E_J\cos(\eta)$ where $\eta$ is conjugate to the number operator $N$, i.e.\ $[N, \eta]=-i$, counting the number of transferred Cooper pairs in units of $2 e$. Due to the usual relation $V_{\rm res}=-(\hbar/2e) \dot{\phi}$ the JJ and the resonator are dynamically coupled.
Alternatively, the JJ part can also be represented in the charge basis as
\begin{equation}
H_J= -\frac{E_J}{2} \sum_N \left( |N\rangle\langle N+1| + |N+1\rangle\langle N|\right)\,,
\end{equation}
where  $|N\rangle$ are eigenstates of the number operator ${N}$. 

The circuit is embedded in fluctuating reservoirs ${\tilde H}_R=H_{R,\rm res}+H_{R, J}$ which are assumed to be statistically independent. $H_{R,\rm res}$ describes high frequency electromagnetic modes constituting a conventional low temperature heat bath with ohmic spectral distribution. $H_{R, J}$ captures the low frequency part of the environment which according to the experiment occurs as voltage noise at the JJ.
 The heat bath  interacts bi-linearly with the phase of the resonator via $H_{R,\rm res}=g_{\rm res}\, \phi\, {\cal E}$ with reservoir force ${\cal E}$ obeying $\langle {\cal E}(t)\rangle_{R,\rm res}=0$ and a second moment fixed by the fluctuation dissipation theorem. The voltage noise can be seen as a fluctuating component to the bias voltage $V\to V+\delta v$ with $\langle \delta v\rangle=0$ and thus couples also to the  charge operator via $H_{I, J}=g_J \, N\, \delta v$.
The interaction between the circuit and the respective environments is sufficiently weak so that a perturbative treatment according to a Born-Markov master equation is applicable.

Now, upon applying the gauge transformation $U_N(t)=\exp[i (\omega_J t + \phi) N]$ with $\omega_J= 2eV/\hbar$, one arrives in the charge representation at
\begin{equation}
\tilde{H}_0'=\frac{(q +2 e {N})^2}{2 C} + \biggl(\frac{\hbar}{2e}\biggr)^2 \frac{1}{2L}\, \phi^2-\frac{E_J}{2} \sum_N \left( {\rm e}^{i \phi}\, {\rm e}^{i \omega_J t}\, |N\rangle\langle N+1| + |N+1\rangle\langle N|\, {\rm e}^{-i \phi}\, {\rm e}^{-i \omega_J t}\right)\, .
\end{equation}
Since the operator $q$ has a continuous spectrum (it describes the continuous displacement of the electronic liquid relative to the ionic background) while that of $N$ is integer, it is convenient to introduce
the charge operator $\tilde{q}=q + 2 e{N}$ which counts charge fluctuations relative to the flow of integer charge quanta through the JJ. The operator $\tilde{q}$ is also conjugate  to $\phi$ and, by slight abuse of notation, will thus be denoted as $q$ again henceforth. Note that the interaction terms with the reservoirs commute with $U_N$.

In a second step, a mapping to a rotating frame is applied via $U_0=\exp(-i \omega_J a^\dagger a t)$ with $a^\dagger$, $a$ denoting the standard annihilation/creation operators for the resonator with $[a,a^\dagger]=1$.
 This then leads to 
 \begin{equation}\label{hrot}
\tilde{H}_{0, \rm rot}'=\hbar\Delta a^\dagger a -\frac{E_J}{2} \sum_N \left( {\rm e}^{i \phi(t)+i \omega_J t}\, |N\rangle\langle N+1| + |N+1\rangle\langle N|\, {\rm e}^{-i \phi(t)-i \omega_J t}\right)\, ,
\end{equation}
where $\Delta=\omega_0-\omega_J$ denotes the detuning. Further, one introduces Heisenberg operators $\phi(t)=\sqrt{\kappa} ( a\, {\rm e}^{-i\omega_J t}+a^\dagger \,{\rm e}^{i\omega_J t})$ with $\kappa=\hbar/2m\omega_0$. Now, a decomposition of the exponentials in (\ref{hrot}) via the Baker-Campbell-Hausdorff formula and representating them in terms of normally ordered Bessel functions $J_1$ of the first kind, one arrives for the time-independent part of $\tilde{H}_{0, \rm rot}'$ at
\begin{equation}
{H}_0=\hbar\Delta a^\dagger a+i\frac{E_J^*}{2} : \left( a^\dagger \, {\rm e}^{i\eta}-  a\, {\rm e}^{-i\eta}\right) \frac{J_1(2\sqrt{\kappa\, n})}{\sqrt{n}}:\, ,
\end{equation}
as given in Eq. (1) of the main text.

We note that in principle the mapping to a rotating frame also affects the interaction with the respective heat baths (see e.g.\cite{verso:2010}), however, the deviations from the standard expressions are typically small and are assumed to be negligible here. As a consequence, based on the total Hamiltonian of circuit and environments and following the standard procedure (second order perturbation theory), the reduced density operator of resonator and JJ obeys at zero temperature a master equation of the form \cite{milburn:1994, carmichaelepaps:2002}
\begin{equation}\label{masterepaps}
\dot{\rho}=-\frac{i}{\hbar}[H_0,\rho]+ \frac{\gamma}{2} (2 a \rho a^\dagger- n \rho- \rho n)+\frac{\gamma_J}{2} (2 {N}\rho\, {N}-{N}^2\, \rho-\rho\, {N}^2)\, .
\end{equation}
  The rate $\gamma\sim O(g_{\rm res}^2)$ describes photon relaxation in the resonator and thus its $Q$-factor $Q=\omega_0/\gamma$, while the rate $\gamma_J\sim O(g_J^2)$ follows from the power of the voltage noise (see below). The dissipators appearing in Eq.~(2) of the main text thus read: ${\cal L}[a,\rho]=(2 a \rho a^\dagger- n \rho- \rho n)$ with $n=a^\dagger a$ as well as ${\cal L}[N,\rho]=(2 {N}\rho\, {N}-{N}^2\, \rho-\rho\, {N}^2)$.

 We now consider the voltage noise in more detail. According to the experimental situation \cite{hofheinzepaps:2011} and in agreement with our simulations based on (\ref{masterepaps}), the voltage noise is negligible for most observables compared to the resonator decoherence due to $\gamma_J/\gamma\approx 0.04\ll 1$. It must only be kept for those quantities which are broadened exclusively due to $\gamma_J$ such as the photon spectrum. Within second order perturbation theory (master equation), one finds $\gamma_J= (4 e^2/\hbar^2)\, D_v(0)$, where 
\begin{equation}
D_v(\omega\to 0)=2\, \int_0^\infty dt \; \Re{\rm e}\{\langle \delta v(t) \delta v(0)\rangle\}
\end{equation}
is the noise power of this low frequency noise. It can be expressed in terms of the low frequency portion $Z_<(\omega)$ of the total environmental impedance as
\begin{eqnarray}
\Re{\rm e}\{\langle \delta v(t) \delta v(0)\rangle\} &=& \frac{\hbar}{\pi} \int_0^\infty d\omega\; \Re{\rm e}\{Z_<(\omega)\}\, \omega \, {\rm coth}(\omega\hbar\beta/2) \cos(\omega t)\nonumber\\
&\approx & \frac{2 k_{\rm B} T}{\pi} \int_0^\infty d\omega\; \Re{\rm e}\{Z_<(\omega)\}\, \cos(\omega t)
\end{eqnarray}
so that
\begin{equation}
D_J(\omega\to 0)\approx 4 k_{\rm B} T\, \Re{\rm e}\{Z_<(\omega\to 0)\}\, .
\end{equation}

\section{Classical equations of motion}

As discussed in the main text, close to the one-photon resonance $\omega_0\approx \omega_J$, the classical steady state orbit may assumed to be of the form $\phi(t)=Z \cos(\omega_J t+\varphi)$ with real-valued amplitude $Z$ and phase $\varphi$. In the rotating frame this then leads for zero detuning to the expression (6) for the classical Hamiltonian, namely, $H_{0,\rm cl}=E_J J_1(Z) \sin(\varphi)$. However, since $Z$ and $\varphi$ are related to the original conjugate variables $p_\phi,\phi$ by a nonlinear transformation, Hamilton's equations of motion must be transformed accordingly. Instead,  it is more convenient to start from the classical limit of the expression for the correlation $\langle a {\rm e}^{-i\eta}\rangle_{\rm st}$ specified in (4) of the main text via the replacement: $2\sqrt{\kappa}a \to Z {\rm e}^{i\varphi}, {\rm e}^{i\eta}\to 1$. Separating in real and imaginary parts then leads to (7) in the main text.

\bibliographystyle{apsrev4-1}
%

\end{widetext}

\end{document}